\documentclass[aps,prl,twocolumn,groupedaddress]{revtex4}
\usepackage[xdvi]{graphicx}
\usepackage{amssymb,amsmath}
\begin{document}

\title{Electronic structure and magnetic properties of FeTe, BiFeO$_3$, 
SrFe$_{12}$O$_{19}$ and SrCoTiFe$_{10}$O$_{19}$ compounds}

\author{G.E.~Grechnev}
\affiliation{B. Verkin Institute for Low Temperature Physics and
Engineering, National Academy of Sciences of Ukraine,
Kharkov 61103, Ukraine}%
\email{grechnev@ilt.kharkov.ua}

\author{A.A. Lyogenkaya}
\affiliation{B. Verkin Institute for Low Temperature Physics and
Engineering, National Academy of Sciences of Ukraine,
Kharkov 61103, Ukraine}%

\author{O.V.~Kotlyar}
\affiliation{B. Verkin Institute for Low Temperature Physics and
Engineering, National Academy of Sciences of Ukraine,
Kharkov 61103, Ukraine}%

\author{A.S.~Panfilov}%
\affiliation{B. Verkin Institute for Low Temperature Physics and
Engineering, National Academy of Sciences of Ukraine,
Kharkov 61103, Ukraine}%

\author{V.P.~Gnezdilov}
\affiliation{B. Verkin Institute for Low Temperature Physics and
Engineering, National Academy of Sciences of Ukraine,
Kharkov 61103, Ukraine}%


\setcounter{page}{1}%

\begin{abstract}
The electronic energy structures and magnetic properties of iron-based compounds with 
group VI elements (FeTe, BiFeO$_3$, SrFe$_{12}$O$_{19}$ and  SrCoTiFe$_{10}$O$_{19}$)
are studied using the density functional theory (DFT) methods.
Manifestations of different types of chemical bonds in magnetism of 
these compounds are studied theoretically.
Calculations of electronic structures of these systems were performed
using the generalized gradient approximation (GGA) for description of 
the exchange and correlation effects within DFT.
For SrFe$_{12}$O$_{19}$ and  SrCoTiFe$_{10}$O$_{19}$ hexaferrites the GGA+$U$ 
method was also employed to deal with strongly correlated 3$d$-electrons.
The calculations have revealed distinctive features of electronic structure of the
investigated iron-based compounds with strongly correlated 3$d$-electrons, 
which can be responsible for their peculiar structural and magnetic properties.
\end{abstract}

\maketitle

\section{Introduction}
The iron-based compounds are usually regarded as strong magnetic materials.
However, recently the new class of iron-based superconductors was 
discovered. 
The representatives of this class are iron-based compounds with group VI elements,
FeS, FeSe and FeTe, which are distinguished by the simplest crystal structure 
among iron-based superconductors, and very different magnetic properties 
\cite{mizuguchi10,martinelli10,viennois10,li09,grechnev12,chen14}.
The characteristic feature of these compounds is interplay of 
antiferromagnetism (AFM) and superconductivity. 

Compounds of iron with another element of group VI, oxygen, are predominantly
belong to ferromagnetic materials. 
There is renewed interest in properties of such iron oxides as
BiFeO$_3$- and $M$Fe$_{12}$O$_{19}$- based systems 
\cite{catalan09,wang09,dionne09,novak05,novak05b,pullar12,liyanage13,vittoria13,feng14}.
BiFeO$_3$ is antiferromagnetic and multiferroic, whereas
$M$Fe$_{12}$O$_{19}$ ($M$= Sr, Ba, Pb) are ferrimagnetic systems
with expected  manifestation of multiferroic properties.
In these compounds a substitution of Fe or the cations 
with other metals can provide a mixed valence state and very
unusual magnetic properties.

Clarification of microscopic mechanisms which determine electric and magnetic 
properties of these oxides and chalcogenides, assumes detailed theoretical 
studying of the electronic structure. 
A number of electronic structure calculations were carried out
for these systems in recent years,
however data on theirs electronic energy structure
are still incomplete and inconsistent.
Also, the electronic states of these systems are regarded as
strongly correlated, and a proper approach has to be
taken for theoretical studies of the electronic structures.

The purpose of this paper is to provide a reliable picture of the 
electronic band structures and corresponding magnetic properties of FeTe, 
BiFeO$_3$,  SrFe$_{12}$O$_{19}$ and  SrCoTiFe$_{10}$O$_{19}$ compounds.
Our study is based on the density functional theory (DFT) methods,
with employing modern extentions to take into account effects of
correlations.

\section{Details of electronic structure calculations}

The self-consistent calculations of electronic structures were carried out by using 
the modified relativistic LMTO method with a full potential 
(FP-LMTO, RSPt implementation \cite{grechnev09,wills10,rspt}) 
and the linearized augmented plane waves method with a full potential 
(FP-LAPW, Elk implementation \cite{elk}).
Exchange and correlation potentials were treated within the generalized 
gradient approximation (GGA \cite{pbe96}) of DFT.
Spin-orbit coupling was included in the self-consistent calculations.

For the employed full potential FP-LMTO and FP-LAPW methods any 
restrictions were not imposed on charge densities or potentials of studied systems, 
that is especially important for anisotropic layered structures of the investigated 
compounds.
In both methods the dual basis set was employed to incorporate two valence 
wave functions with the same orbital quantum number (e.g. 3$p$ and
4$p$ functions for Fe).
Electronic structure calculations for compounds were carried out for
experimental crystal lattice parameters. 

\section{Electronic structure and magnetic properties of F${{\rm \bf e}}$T${{\rm \bf e}}$}

At temperatures above 70 K FeTe compound possesses the tetragonal PbO-type  
crystal structure (space group $P4/nmm$).
With decreasing temperature, at $T\simeq 70$ K, FeTe demonstrates 
a first-order structural phase transition from tetragonal to monoclinic structure. 
This transition is accompanied by bicollinear AFM ordering, as determined by means 
of x-ray and neutron diffraction studies \cite{mizuguchi10,martinelli10,viennois10,li09}, 
and we used the experimental data on structural parameters $a$, $b$ and $c$ in our calculations.
The angle between axes was taken as 90 $^{\circ}$  
(instead of 89,2 $^{\circ}$ for the actual structure with small monoclinic distortion).

\begin{figure}[]
\begin{center}
\includegraphics*[trim=0mm 0mm 0mm 0mm,scale=0.5]{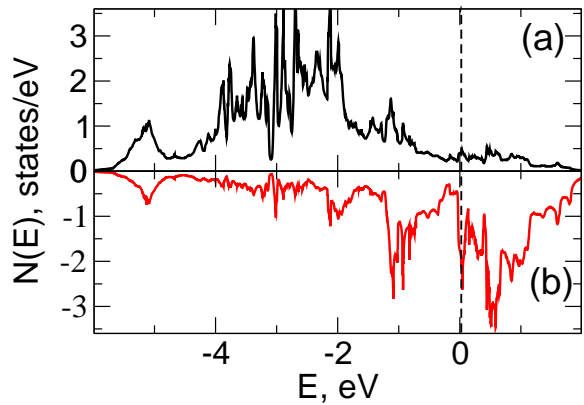}
\caption{\label{fig1_fete_dos} 
(Color online) Spin-polarized densities of states $N(E)$ of FeTe compound for 
the double-stripe AFM phase (per formula unit): positive ((a), black) and 
negative ((b), red) values of DOS correspond to majority and minority spin states, 
respectively. 
The position of Fermi level ($E_F=0$) is marked by a vertical dashed line.
}
\end{center}
\end{figure}

The calculations of electronic structure were carried out in the GGA-approximation  
for magnetic phases of FeTe (ferromagnetic, collinear AFM, bicollinear AFM) and it was
demonstrated, that minimum of total energy is found for the bicollinear 
(or ``double stripe'', DS) AFM phase, which appears to be a ground state of FeTe compound.
The calculated for the ground state spin-polarized densities of electronic states (DOS) 
are presented in Fig. \ref{fig1_fete_dos}.
As can be seen in Fig. \ref{fig1_fete_dos} (b), in the DS AFM phase of FeTe 
the Fermi level is located in very close proximity ($\sim$0.1 eV) of 
the sharp peak of DOS.

\begin{figure}[]
\begin{center}
\includegraphics*[trim=0mm 0mm 0mm 0mm,angle=90,scale=0.18]{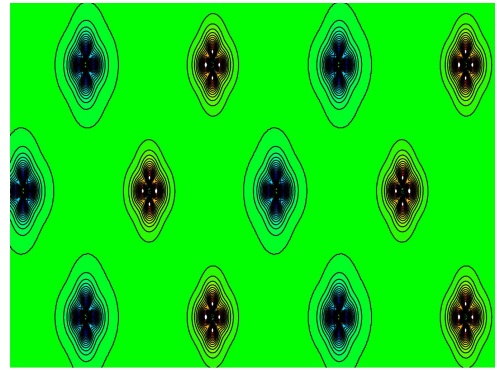}
\caption{\label{fig2_fete_sd}
(Color online) Calculated spin density contours in the (001) plane
for the double-stripe antiferromagnetic phase of FeTe.
(For clarification of this color figure, 
the reader is referred to the web version of the paper.)
}
\end{center}
\end{figure}

The spin-polarized splitting of DOS $N(E)$ (see Fig. \ref{fig1_fete_dos})
provides formation of magnetic moments at Fe sites for DS AFM phase of FeTe.
The obtained in our calculations value of the magnetic moment 
$M_{\rm Fe}\cong 2.37 \mu_{\rm B}$ is in agreement with the results of 
neutron diffraction studies
($M_{\rm Fe}^{\rm exp}=2.26\div 2.54 \mu_{\rm B}$ \cite{martinelli10,li09}). 
Such good agreement with experiment confirms the itinerant nature of magnetic
moments in FeTe and adequacy of DFT--GGA approximation which we used in the present 
calculations of magnetic characteristics.
The calculated spin density contours in the (001) plane for the AFM phase
of FeTe compound are presented in Fig. \ref{fig2_fete_sd}.
The double stripe structure of the ground state AFM phase is
clearly seen in this figure.

\begin{figure}[]
\begin{center}
\includegraphics*[trim=0mm 0mm 0mm 0mm,angle=0,scale=0.35]{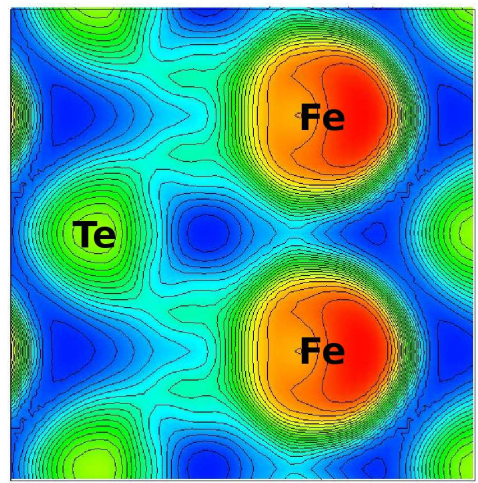}
\caption{\label{fig3_fete_cd}
(Color online) Calculated charge density contours in the (100) 
plane for FeTe compound in DS AFM phase.
(For clarification of this color figure, 
the reader is referred to the web version of the paper.)
}
\end{center}
\end{figure}

The calculated charge density contours in the vertical (100) plane of unit cell of FeTe 
compound are presented in Fig. \ref{fig3_fete_cd}. 
It demonstrates a substantial anisotropy of the charge density distribution between
iron atom and neighbouring atoms of tellurium.
It is plausible to assume, that formation of directed covalent bonds between 
the neighbouring Fe and Te atoms, seen in Fig. \ref{fig3_fete_cd}, is caused 
by hybridization of $d$-states of iron and $p$-states of tellurium. 

\begin{figure}[h]
\begin{center}
\includegraphics*[scale=0.5]{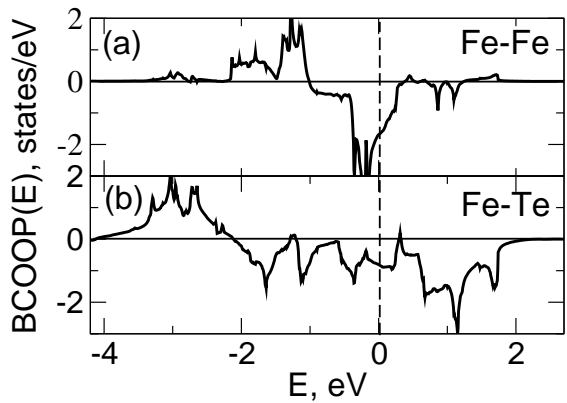}
\caption{\label{fig4_fete_coop} Balanced crystal orbital overlap 
populations BCOOP($E$) of FeTe compound in the DS antiferromagnetic phase 
for the pairs of nearest atoms in the unit cell:
(a) Fe---Fe; (b) Fe---Te.
The position of Fermi level ($E_F=0$) is indicated by a vertical dashed line.}
\end{center}
\end{figure}

For a more detailed study of the chemical bonds in FeTe compound, we have calculated 
the crystal orbital overlap populations (BCOOP, \cite{agrechnev03}) by using 
the FP-LMTO method \cite{wills10}. 
The calculated BCOOP($E$) distributions (see Fig. \ref{fig4_fete_coop}) represent 
generalization for solids of the crystal orbital overlap population (COOP) molecular 
characteristics, known in quantum chemistry. 
The values of BCOOP($E$) depend on the energy of the electronic states in
the valence band and are positive for bonding orbitals and
negative for antibonding orbitals, as well as in the case of ionic
bonding \cite{agrechnev03}.
The metallic type bonding is characterized by negative values of BCOOP($E$) 
at energies close to the Fermi level.

According to results of BCOOP($E$) calculations,  
the overlap of Fe--Te orbitals gives bonding states closer to the bottom of valence band 
(positive BCOOP($E$) in the range  $-4\div -2$ eV, see Fig. \ref{fig4_fete_coop} (b)).
The bonding states are also formed at Fe--Fe orbital overlapping 
in the energy range $-3\div -1$ eV (Fig. \ref{fig4_fete_coop} (a)).
In the vicinity of the Fermi energy, in the range $-1\div 1$ eV, for Fe--Te and Fe--Fe bonds 
the negative values of BCOOP($E$) were obtained (Fig. \ref{fig4_fete_coop} (a) and (b)), 
and this corresponds to a metallic type of bonding.
Basically, the calculated distribution of electronic density in Fig. \ref{fig3_fete_cd} 
definitely has features of metallic bonding, whereas distinct covalent bonds are 
formed between Fe and Te atoms.

\begin{figure}[h]
\begin{center}
\includegraphics*[trim=20mm 0mm 20mm 0mm,scale=0.3]{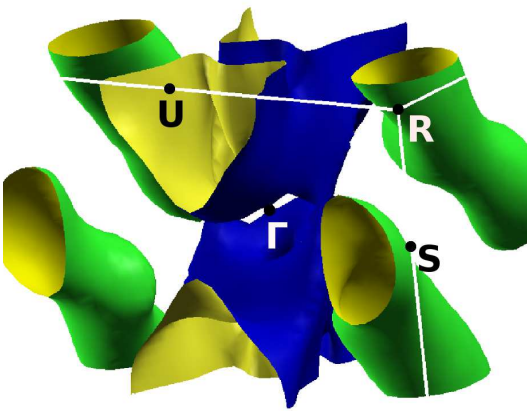}
\caption{\label{fig5_fete_fs}
(Color online) Calculated Fermi surface of FeTe compound in the bicollinear AFM phase.
(For clarification of this color figure, 
the reader is referred to the web version of the paper.)
}
\end{center}
\end{figure}

In this work the Fermi surface (FS) of the low-temperature DS AFM phase of FeTe 
was calculated for the first time, and it is presented in Fig. \ref{fig5_fete_fs}.
The complicated shape of this FS differs drastically from that of paramagnetic FeTe,
which was calculated earlier \cite{subedi08} and represents 
compensating electron and hole cylinders.
This radical FS reconstruction at the AFM transition can be the origin of 
the sign reversal of the Hall coefficient which has been observed in FeTe \cite{liu11}.
Due to the multiband electronic structure of FeTe, the AFM transition
presumably initiates a reversal of the balance between electron and hole
contributions to the Hall coefficient.

\section{Electronic structure and magnetic properties of  B${{\rm \bf i}}$F${{\rm \bf e}}$O$_3$}

The compound BiFeO$_3$ is a typical representative of substances known as 
magnetoelectrics or multiferronics, in which magnetic and electric orderings coexist.
Bismuth ferrite BiFeO$_3$ has a relatively simple crystalline structure of
perovskite type and is of interest as a model object for theoretical studies of 
magnetic and electrical properties.
In the last decade nonempirical calculations of the electronic structure were performed 
for different crystalline modifications of bismuth ferrite, and the
importance of including Coulomb correlations in the LSDA+U or GGA+U approximations for 
a correct description of the electronic structure and ferroelectric properties of
BiFeO$_3$ was indicated \cite{neaton05,clark07,mcleod10}.
On the other hand, the possibility of describing the electronic structure and magnetism 
of BiFeO$_3$ within the density functional theory in the generalized gradient approximation 
has not been consistently and comprehensively investigated.
Here we have calculated electronic structure and magnetic properties of BiFeO$_3$ 
in the DFT--GGA approximation, without including the Coulomb parameter U, 
with the FP-LAPW method including spin-orbital coupling \cite{elk}.

\begin{figure}[]
\begin{center}
\includegraphics*[scale=0.5]{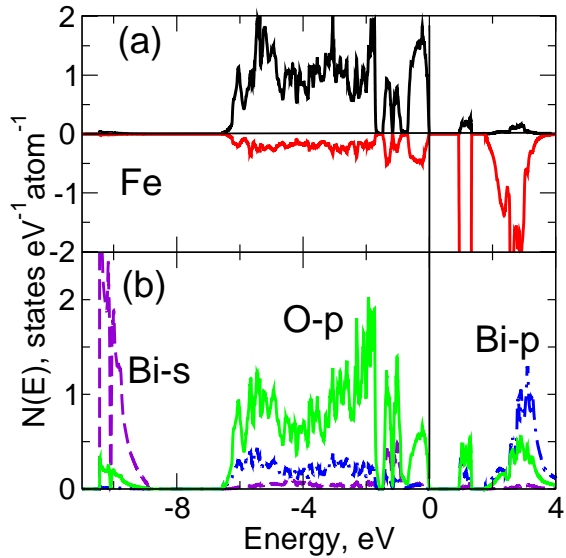}
\caption{\label{fig6_bfo_dos}(Color online)
Partial densities of electronic states $N(E)$ of BiFeO$_3$ 
in the $G$-type antiferromagnetic phase: 
(a) spin-polarized densities of states per Fe atom: 
positive (black) and negative (red) values of DOS correspond to majority 
and minority spin states, respectively; 
(b) densities of states of $s$- and $p$-type per Bi atom: dashed (violet) 
and dashed-dotted (blue) curves, respectively; 
density of states per oxygen atom: solid green curve. 
The Fermi level ($E_F=0$) is indicated by a vertical line.
(For clarification of the references to color in this ﬁgure caption, 
the reader is referred to the web version of this paper.)
}
\end{center}
\end{figure}

At low temperatures BiFeO$_3$ has a rhombohedral distorted perovskite crystal structure 
(space group $R3c$). 
The electronic structure was calculated for the experimental values  of crystal lattice parameters,
assuming the collinear $G$-type antiferromagnetic ordering \cite{neaton05}.
The rhombohedral angle of the $R3c$ structure was fixed at its experimental value 
of $\alpha_R=59.35^{\circ}$.
The calculated partial electronic densities of states (DOS) $N(E)$ for
BiFeO$_3$ in the $G$-type antiferromagnetic phase are shown in
Fig. \ref{fig6_bfo_dos}. 
This figure shows that the  $G$-type AFM phase is characterized by an insulating state with 
an energy gap of about 1 eV, in good agreement with emission and absorption 
spectroscopy data \cite{mcleod10}.

In Fig.  \ref{fig6_bfo_dos} (b) one can see a narrow band at energies about $-$10 eV
originating from 6$s$-states of Bi and hybridized with 2$p$-states of oxygen. 
On the other hand, the 6$p$-states of Bi are substantially higher in energy. 
It can be expected that this narrow band, which is predominantly of 6$s$-states of Bi,
corresponds to so-called stoichiometrically active ``lone pair'' electrons, 
which are assumed to be responsible for polarization in ferroelectrics 
based on bismuth and lead \cite{catalan09,wang09}.
Due to hybridization with 2$p$-states of oxygen, this 6$s$ ``lone pair'' can no 
longer have a purely spherical spatial charge distribution, but can acquire 
a component in the form of a ``lobe'', which is characteristic for $p$-orbitals. 
Then, the Bi($s$)--O($p$) hybridization leads to a noticeable spatial anisotropy 
in the charge density distribution.
Thus, hybridization of $sp$-orbitals of Bi with 2$p$-orbitals of oxygen leads to 
asymmetric charge transfer in the Bi--O bonds which evidently facilitates the 
development of ferroelectricity.

\begin{figure}[]
\begin{center}
\includegraphics*[scale=0.5]{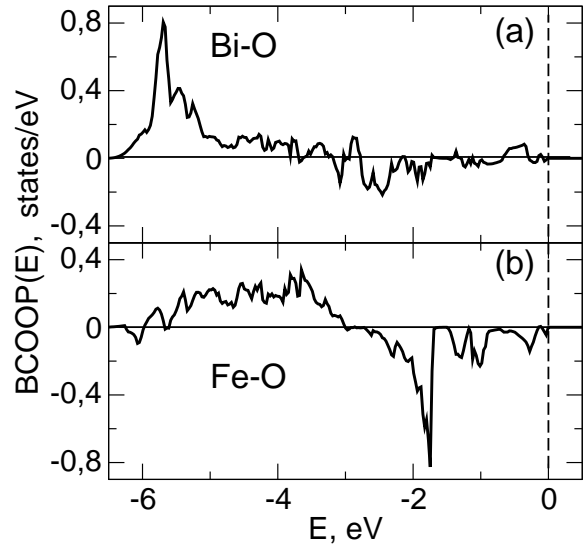}
\caption{\label{fig7_bfo_coop}
Crystal orbital overlap populations BCOOP($E$) of BiFeO$_3$ in the 
G-type antiferromagnetic phase for the pairs of nearest atoms in the
unit cell: Bi--O (a), Fe--O (b). The top of the valence band ($E=0$)
is indicated by a vertical dashed line.}
\end{center}
\end{figure}

For a more detailed study of the chemical bonds in BiFeO$_3$, we have calculated 
the crystal orbital overlap populations (BCOOP, \cite{agrechnev03}) by using 
the FP-LMTO method \cite{wills10}. 
The calculated BCOOP($E$) distributions (see Fig. \ref{fig7_bfo_coop}) represent generalization of 
the crystal orbital overlap population (COOP) characteristics from quantum chemistry for solids. 
The values of BCOOP($E$) depend on the energy of the electronic states in
the valence band and are positive for bonding orbitals and
negative for antibonding orbitals, as well as in the case of ionic
bonding \cite{agrechnev03}.

According to the BCOOP(E) calculations, the most distinct bonding states in 
the valence band of BiFeO$_3$ are formed in the FeO$_6$ octahedra upon hybridization 
of $d$-states of iron with $p$-states of oxygen (Fig. \ref{fig7_bfo_coop} (b)). 
This is consistent with the exact coincidence in energy of the dominant regions
of the partial densities of the $d$-states of iron (Fig. \ref{fig6_bfo_dos} (a)) and
$p$-states of oxygen  (Fig.  \ref{fig6_bfo_dos} (b)) in the valence band of  BiFeO$_3$.

Overlap of the Bi--O orbitals yields bonding states near the bottom of the valence 
band (positive BCOOP($E$) in Fig.  \ref{fig7_bfo_coop} (a)).
It should be noted that the contribution of the $p$-states of Bi in the 
valence band is substantially depleted (Fig.  \ref{fig6_bfo_dos} (b)) due to 
the charge transfer from Bi to neighbouring Fe and O atoms. 
This can be the origin of the ionic bonding for Bi--O orbitals 
in the upper part of the valence band (see Fig. \ref{fig7_bfo_coop} (a)).

On the other hand, the ionic bonding is even more pronounced for Fe--O bonds 
near the top of the valence band (Fig. \ref{fig7_bfo_coop} (b)). 
The corresponding charge transfer in Fe--O pairs provides the 
distortion of the FeO$_6$ octahedra in BiFeO$_3$ \cite{catalan09,wang09}.
In fact, our calculations indicate a predominantly ionic character of
the chemical bonds in BiFeO$_3$. 
Hybridization of Fe($d$)- and O($p$)-states also produces a partially covalent 
component of the Fe--O bond. 

The antiferromagnetic ordering in BiFeO$_3$ provides the spin-split density of states $N(E)$,
and the magnetic moments are predominantly formed at iron ions (see. Fig.  \ref{fig6_bfo_dos} (a)).
The calculated value of $M_{\rm Fe}\cong 3.7 \mu_{\rm B}$ agrees
with data from neutron diffraction studies ($M_{\rm Fe}\cong 3.75 \mu_{\rm B}$ \cite{sosnowska02}).
This good agreement with experiment supports the reliability of the DFT-GGA approach used 
here, which can be applied for studying magnetic properties of BiFeO$_3$ under pressure and doping.
It is also revealed, that a part of the total magnetic moment of the unit cell 
is distributed among neighbouring oxygen ions (up to 0.2 $\mu_{\rm B}$ per O ion).

\section{Electronic structure and magnetic properties of 
S${{\rm \bf r}}$F${{\rm \bf e}}_{12}$O$_{19}$ and 
S${{\rm \bf r}}$C${{\rm \bf o}}$T${{\rm \bf i}}$F${{\rm \bf e}}_{10}$O$_{19}$}

Hexaferrites SrFe$_{12}$O$_{19}$, BaFe$_{12}$O$_{19}$, and PbFe$_{12}$O$_{19}$
are known as strong permanent magnets, and have been intensively studied for 
the last decade due to their intriguing magnetic and chemical properties.
Many efforts were undertaken to tune magnetic properties of
hexaferrites by substituting  Fe$^{3+}$, Ba$^{2+}$ and Sr$^{2+}$ cations 
with other transition metal or rare earth ions at various lattice sites
(see Refs. \cite{novak05,liyanage13,vittoria13,feng14} and references therein). 
Recently, the SrFe$_{12}$O$_{19}$ hexaferrite was found to become
magnetoelectric at room temperature by substitution of Fe with 
Ti and Co \cite{vittoria13,feng14}.
The aim of the present investigation was, firstly, to find a reliable approach
of DFT to describe properly electronic structure and magnetism of SrFe$_{12}$O$_{19}$, 
and then, using this approach, to explore electronic structure and 
magnetic properties of SrFe$_{12}$O$_{19}$-based systems where 
Fe is substituted with Ti and Co atoms.

Compound SrFe$_{12}$O$_{19}$ possess the hexagonal crystal structure
which belongs to the $P6_{3}/mmc$ space symmetry group \cite{novak05}.
The double formula unit cell contains 64 atomic sites distributed among 
11 inequivalent Wyckoff positions: 2$d$ (Sr), 2$a$ (Fe), 2$b$ (Fe), 4$f_1$ (Fe), 
4$f_2$ (Fe), 12$k$ (Fe), 4$e$ (O), 4$f$ (O), 6$h$ (O), 12$k_1$ (O), 12$k_2$ (O).
According to the previous studies \cite{dionne09,novak05,pullar12,liyanage13}, the stable 
spin configuration is assumed to be ferrimagnetic with Fe at the 4$f_1$ and 4$f_2$ 
sites having the magnetic moment anti-parallel to the rest of the Fe cations.

\begin{figure}[]
\begin{center}
\includegraphics*[scale=0.5]{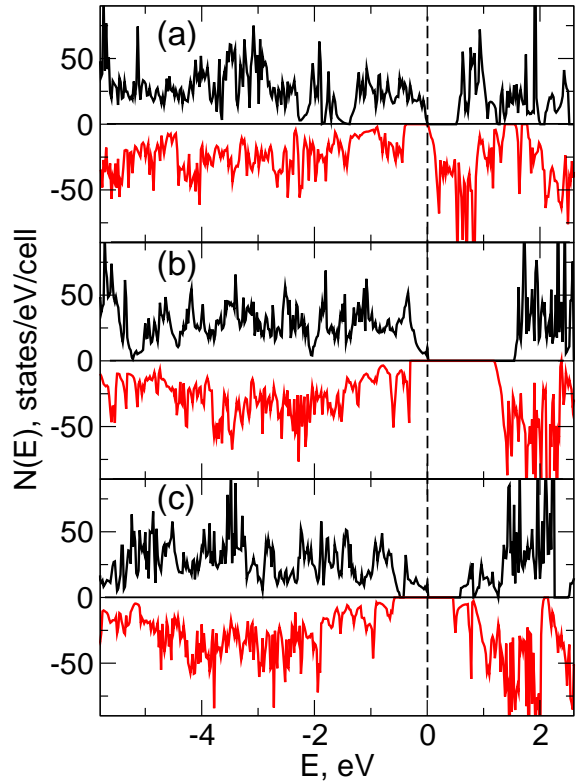}
\caption{\label{fig8_sfo_dos}(Color online)
Total spin-polarized densities of electronic states of 
strontium hexaferrites:
(a)  SrFe$_{12}$O$_{19}$, GGA;
(b)  SrFe$_{12}$O$_{19}$, GGA+U;
(c)  SrCoTiFe$_{10}$O$_{19}$, GGA+U.
Positive (black) and negative (red) values of DOS correspond to majority 
and minority spin states, respectively.
The Fermi level ($E_F=0$) is indicated by a vertical dashed line.
}
\end{center}
\end{figure}

It has been established, that LSDA approach has provided
incorrectly metallic ground state of 
$M$Fe$_{12}$O$_{19}$ hexaferrites \cite{novak05,novak05b}.
Therefore for SrFe$_{12}$O$_{19}$ we firstly used DFT GGA approach 
\cite{pbe96}, which usually gives more reliable picture of 
the ground state for systems with 3$d$-electrons. 
The GGA+$U$ approach was also employed within FP-LAPW method
\cite{elk} in line with Ref. \cite{liechtenstein} for LSDA+$U$, 
but with the GGA instead of LSDA exchange-correlation potential.
It is believed that this approach provides better description 
of electronic structure for localized electrons in 
the 3$d$-transition metal oxides. 
The on-site Coulomb repulsion energy $U$ was taken close to previous 
{\em ab initio} estimates \cite{novak05,liyanage13}, $U$=4 eV.
The exchange parameter $J$ was assumed to be close to its
atomic value $J\simeq 1$ eV  \cite{liechtenstein}.

The results of GGA and GGA+$U$-based calculations of DOS for SrFe$_{12}$O$_{19}$ 
compound are presented in Fig.~\ref{fig8_sfo_dos}, (a) and (b), respectively.
As can be seen in Fig.~\ref{fig8_sfo_dos} (a), the GGA FP-LAPW calculations 
for the spin-polarized ferrimagnetic state have provided 
the almost zero gap semiconductor.
The energy gaps are enlarged in both majority and minority spin channels,
which are moved apart and almost overlaped. 
For the GGA+$U$ calculations the insulating ground state was obtained
in agreement with experiment (see Refs. \cite{dionne09,pullar12}) 
with the energy gap $E_G\simeq 1.2$ eV, 
as is seen in Fig.~\ref{fig8_sfo_dos} (b).

\begin{table}[]
\caption{Local magnetic moments of ions in different Wyckoff sites of SrFe$_{12}$O$_{19}$
and SrCoTiFe$_{10}$O$_{19}$ calculated using GGA and/or GGA+$U$ functionals. 
Magnetic moments $M$ are given in $\mu_B$.}
\label{tab1}
\begin{center}
\begin{tabular}{|c|c|c|c|c|c|}
\hline
 Wyckoff & \multicolumn{3}{|c|}{SrFe$_{12}$O$_{19}$} & \multicolumn{2}{|c|}{SrCoTiFe$_{10}$O$_{19}$}\\
\cline{2-6}
position & atom  & $M$, GGA &  $M$, GGA+U &   atom  & $M$, GGA+U        \\
\hline
2$d$  & Sr &-0.016 &-0.018 & Sr & -0.017 \\
2$a$  & Fe & 3.748 & 4.052 & Ti  & 0.215 \\
2$b$  & Fe & 3.583 & 3.934 & Co  & 2.604  \\
4$f_1$& Fe &-3.490 &-3.907 & Fe  &-3.910  \\
4$f_2$& Fe &-3.258 &-3.922 & Fe  &-3.911   \\
12$k$ & Fe & 3.768 & 4.066 & Fe  & 4.071  \\
4$e$  & O  & 0.492 & 0.492 & O  &  0.497 \\
4$f$  & O  & 0.142 & 0.123 & O  &  0.110 \\
6$h$  & O  & 0.100 & 0.051 & O  &  0.069 \\
12$k_1$ & O  & 0.123 & 0.114 & O  &  0.007 \\
12$k_2$ & O  & 0.217 & 0.218 & O  &  0.220 \\
\hline
\end{tabular}
\end{center}
\end{table}

The magnetic moments for all inequivalent atoms in the unit cell
of SrFe$_{12}$O$_{19}$, which were calculated with both GGA and GGA+$U$
approaches, are listed in Table \ref{tab1}. 
For the insulating GGA+$U$ case the sum of all magnetic moments  per unit
cell equals to 40 $\mu_B$, which formally corresponds to 5 $\mu_B$ per Fe ion 
in the adopted ferrimagnetic structure.
In fact, however, the theoretical value of 5 $\mu_B$ corresponding to the 
free Fe$^{3+}$ ion is reduced in a crystal field environment, and a part 
of the total magnetic moment is transferred to neighbouring oxygen ions,
as is seen in  Table \ref{tab1}. 
Actually, the Fe ions in SrFe$_{12}$O$_{19}$ have magnetic moments close 
to 4 $\mu_B$, as it comes from GGA+$U$ calculations.

The electronic structure of SrCoTiFe$_{10}$O$_{19}$ compound 
was calculated using the GGA+$U$ method.
Following results of previous studies for 3$d$-transition metal oxides 
\cite{dionne09,feng14}, we used $U=4$ eV for Fe and $U=3$ eV for Co as 
the appropriate values for our GGA+$U$ calculations.
Based on total energy calculations, we conclude that in the 
ground state configuration Ti and Co ions preferentially occupy
2$a$ and 2$b$ positions, respectively.
It should be noted, that configuration with Ti and Co ions 
in 12$k$ and 4$f_1$ positions, respectively, corresponds
to slightly higher total energy, and also requires consideration for 
the recently synthesized SrCo$_2$Ti$_2$Fe$_{8}$O$_{19}$ 
multiferroic compound \citep{feng14}.

Our GGA+$U$ calculations have revealed the insulating ground state
for SrCo$_2$Ti$_2$Fe$_{8}$O$_{19}$ with the energy gap $E_G\simeq 0.5$ eV, 
as is seen in Fig.~\ref{fig8_sfo_dos} (c).
The magnetic moments on Co atoms are found to be 2.6 $\mu_B$
(see Table \ref{tab1}), which is lower than that corresponding to
a high spin state of Co$^{2+}$ ion.
As in the case of Fe$^{3+}$ ions, the high value of the moment 
corresponding to free Co$^{2+}$ ion is reduced in the crystal field, 
and the total magnetic moment is partly distributed among 
neighbouring oxygen ions, as is seen in  Table \ref{tab1}. 
The sum of all magnetic moments per unit cell of SrCo$_2$Ti$_2$Fe$_{8}$O$_{19}$
equals to 28 $\mu_B$. 

\section{Conclusions}

Results of the present GGA DFT calculations indicate predominantly metallic 
character of chemical bonding in FeTe compound, with partially covalent 
components of Fe--Te, Te--Te and Fe--Fe bonds.
In fact, a hybridization of $d$-states of iron with $p$-states of tellurium
results in the pronounced anisotropy of spatial distribution of
charge density in area between the planes of Fe and Te, and also
in the Te$\to$Fe charge transfer.
The presence of covalent bonding presumably promotes stabilization of monoclinic 
structural distortions of the tetragonal phase of FeTe at low temperatures.

Results of the present calculations affirm that magnetic properties of FeTe
are well described within the approach of itinerant electrons and GGA DFT.
In particular, it is established that the bicollinear AFM phase has the lowest 
total energy, being the ground state of FeTe compound.
The calculated value of magnetic moment for the bicollinear AFM phase
($M_{\rm Fe}\cong 2.4 \mu_{\rm B}$) is very close to results 
of neutron diffraction experiments.

On the other hand, in BiFeO$_3$ compound the GGA DFT calculations revealed that 
the chemical bonding is predominantly of an ionic character with partial covalent 
components of the Bi--O and Fe--O bonds. 
For purely ionic bonding the spatial distribution of the charge would be purely 
spherical near the Bi, Fe, and O ions with keeping of an ideal perovskite structure. 
A covalent bonding facilitates stabilization of the structural distortions, 
which favours formation of a ferroelectric phase in BiFeO$_3$. 
In particular, hybridization of $s,p$-states of Bi with 2$p$-states of oxygen leads 
to a distinct spatial anisotropy of the charge density distribution and to an 
asymmetry in charge transfer in Bi--O bonds, which presumably causes ferroelectric 
polarization in BiFeO$_3$. 
On the other hand, hybridization of Fe($d$)- and O($p$)-states leads to substantial spin 
polarization in the AFM $G$-phase and to distortion and turning of the FeO$_6$ octahedra. 
This hybridization reduces the value of magnetic moment of free Fe$^{3+}$ ion to
$M_{\rm Fe}\cong 3.7 \mu_{\rm B}$, and it is also responsible for spin polarization 
of electronic states of oxygen ions, providing the noticeable magnetic moment 
$M_O\simeq 0.1 \mu_B$ for each O ion. 

The electronic structures of hexagonal strontium ferrites SrFe$_{12}$O$_{19}$ and 
SrCoTiFe$_{10}$O$_{19}$ were also calculated based on the generalized gradient 
approximation of DFT. 
However, the GGA+$U$ method has been employed to improve the description of localized 
3$d$-electrons, and to reproduce the insulating ground states of these compounds,
with the energy gaps about 1 eV.
Resultantly, the calculated magnetic moments of 3$d$-atoms are notably lower than 
the high spin values of the moments of free Fe$^{3+}$ and Co$^{2+}$ ions.
In fact these moments are reduced in the crystal field environment, 
and the total magnetic moment is partly distributed among neighbouring oxygen ions.
The total magnetic moment of ferrimagnetic SrCoTiFe$_{10}$O$_{19}$ depends on the 
particular sites occupancy with substitution of Fe$^{3+}$ ions by Co$^{2+}$ and Ti$^{4+}$. 

The results of the present work indicate that theoretical analysis of electronic structures
of iron-based compounds with group VI elements (chalcogenides and oxygen) by means of GGA and
GGA+$U$ DFT calculations provides avenue to explore peculiar magnetic properties of different 
systems with strongly correlated electrons.
This includes superconductors with AFM ordering, various multiferroics, and strong ferrimagnets.

\begin{acknowledgments}
This work was supported by the Russian-Ukrainian RFBR-NASU
project 78-02-14, and was performed using computational facilities of 
grid-cluster ILTPE — B. Verkin Institute for Low Temperature Physics 
and Engineering of the National Academy of Sciences of Ukraine.
\end{acknowledgments}

\end{document}